\documentclass[11pt,a4paper]{article}
\usepackage{jheppub}
\usepackage{amsmath}
\usepackage{amssymb}
\usepackage{color, colortbl}

\newcommand{\bP}{{\bf P}}

\newcommand{\fldZ}{\mathcal{Z}}

\newcommand{\fldD}{\mathcal{D}}

\newcommand{\transcendentalitylevel}{transcendentality}
\newcommand{\transcendentalitylevels}{transcendentalities}

\newcommand{\alg}[1]{\mathfrak{#1}}

\newcommand{\gym}{g\indups{YM}}
\newcommand{\indups}[1]{_{\mathrm{\scriptscriptstyle #1}}}
\newcommand{\gaba}{\gamma\inddowns{ABA}}
\newcommand{\inddowns}[1]{^{\mathrm{\scriptscriptstyle #1}}}

\newcommand{\veci}{|\,\vec{i}\,|}

\newcommand{\ABA}{\mbox{\tiny ABA}}
\newcommand{\Wrap}{\mathrm{wrap}}

\newcommand{\Tr}{{\rm Tr \,}}
\newcommand{\Op}{\mathcal{O}}

\newcommand{\cN}{\mathcal{N}}
\newcommand{\cC}{\mathcal{C}}

\def\HBS{\mathbb S}
\def\RRS{\mathbb{R}}
\def\HS{S}
\def\reciP{\mathcal{P}}

\def\cP{\mathcal{P}}
\def\cT{\mathcal{T}}

\newcommand{\beq}{\begin{equation}}
\newcommand{\eeq}{\end{equation}}
\newcommand{\beqa}{\begin{eqnarray}}
\newcommand{\eeqa}{\end{eqnarray}}

\newcommand{\h}{{\mathrm{h}}}

\newcommand{\sfrac}[2]{{\textstyle\frac{#1}{#2}}}

\newcommand{\M}{M}

\def\z#1{{{\zeta_#1}}}
\def\zp#1#2{{{\zeta_#1^#2}}}
\def\zp#1#2{{{\zeta_#1^#2}}}

\title{Twist-2 at seven loops in planar $\cN=4$ SYM theory: Full result and analytic properties}

\author[a,b]{Christian Marboe}
\author[c,b]{~~~Vitaly Velizhanin}

\affiliation[a]{School of Mathematics, Trinity College Dublin, College
  Green, Dublin 2, Ireland.}

\affiliation[b]{
Institut f{\"u}r  Mathematik und Institut f{\"u}r Physik,
Humboldt-Universit{\"a}t zu Berlin,
IRIS Adlershof, Zum Gro\ss{}en Windkanal 6,
12489 Berlin, Germany.
}

\affiliation[c]{
Theoretical Physics Division,
NRC "Kurchatov Institute",
Petersburg Nuclear Physics Institute,
Orlova Roscha, Gatchina,
188300 St.~Petersburg, Russia.}

\emailAdd{marboec@tcd.ie}

\emailAdd{velizh@thd.pnpi.spb.ru}

\abstract{
The anomalous dimension of twist-2 operators of arbitrary spin in planar $\cN=4$ SYM theory is found at seven loops by using the quantum spectral curve to compute values at fixed spin, and reconstructing the general result using the LLL-algorithm together with modular arithmetic. The result of the analytic continuation to negative spin is presented, and its relation with the recently computed correction to the BFKL and double-logarithmic equation is discussed.
}

\begin{document}

\maketitle

\section{Introduction}
\label{sec:intro}

In our previous work~\cite{Marboe:2014sya} we found the general form of the six-loop anomalous dimension of twist-2 operators in planar $\cN=4$ SYM theory by exploiting the {\texttt{MATHEMATICA} realisation~\cite{Marboe:2014gma} of the solution of the weak coupling solution of the quantum spectral curve (QSC)~\cite{Gromov:2013pga,Gromov:2014caa} to compute values at fixed spin, and using the LLL-algorithm~\cite{Lenstra:1982}\footnote{We used the \texttt{fplll}-program~\cite{fplll}, a \texttt{C++} realisation of the LLL-algorithm.}
to reconstruct the full result. This result, being analytically continued, provided us with information about the generalised double-logarithmic equation~\cite{Gorshkov:1966qd,Kirschner:1983di,Velizhanin:2011pb} and the Balitsky-Fadin-Kuraev-Lipatov (BFKL) equation~\cite{Lipatov:1976zz,Kuraev:1977fs,Balitsky:1978ic}. The last information allowed a reconstruction of the eigenvalue of the kernel of the BFKL equation in the next-to-next-to-leading logarithmic approximation (NNLLA), which is very interesting for the study of the corrections to the BFKL equation. Using some guesses, this reconstruction was done by one of the authors~\cite{Velizhanin:2015xsa}. A result for a similar quantity was obtained directly by solving the QSC perturbatively as a double scaling expansion~\cite{Gromov:2015vua}. The seven loop contribution to the anomalous dimension of twist-2 operators will allow us to check this new result for the BFKL equation, and having in hand all necessary tools we decided to extend our previous computations to one loop more.  

The usage of the very powerful \texttt{C++} realisation~\cite{fplll} of the LLL-algorithm~\cite{Lenstra:1982} in the form of the \texttt{fplll}-program, and the knowledge of the general form of the full six-loop anomalous dimension, makes it possible to reconstruct the full seven loop anomalous dimension directly from a set of fixed values, i.e. without subdividing into the parts coming from the asymptotic Bethe ansatz (ABA) and the wrapping correction as we did in our computation of the six loop result. 
Indeed, we were able to construct all parts of the result that are proportional to zeta-values without this subdivision. However, for the rational part of the result we were unable to produce enough data points from the QSC method\footnote{The LLL-algorithm demands a lot of computer time, which can be resolved, in principle, with the parallelisation of the computation, which is not available at this moment, as usual applications of the LLL-algorithm do not involve such huge matrices and numbers as in our case.}, and thus the subdivision was necessary in this case. As the basis for the rational part coming from the ABA is very large, we numerically compute the same number of values as the number of harmonic sums in the basis with a very high precision and solve the obtained system of linear equations. To compute the rational part of the wrapping correction, i.e. the difference between the full result and the ABA result, we used a slightly modified version of the \texttt{MATHEMATICA} realisation~\cite{Marboe:2014sya} of the QSC-method, which computes only the rational part, and this allows to considerably extend our dataset as one of the most time-consuming parts of the algorithm is the Laurent expansion of $\eta$-functions, which can be significantly simplified when zeta-values are ignored.

In section~\ref{sec:ABA} we describe the computation of the rational part of the contribution coming from the ABA to the seven-loop anomalous dimension of twist-2 operators.
In section~\ref{sec:finiteS} we briefly describe the perturbative solution of the quantum spectral curve used to compute the seven-loop anomalous dimension at fixed spin, and the modifications that makes it possible to work with partial results.
In section~\ref{sec:L7} we reconstruct the general form of the seven-loop anomalous dimension from the fixed values. 
In section~\ref{sec:weak} we provide the constraints which are used to verify the obtained result, together with the description of their origin.

\section{The seven-loop anomalous dimension from Bethe ansatz}\label{sec:ABA}

In this section we briefly give formulas, which can be used for the computation of the ABA part of the anomalous dimension of twist-2 operators in planar $\cN =4$ SYM theory at seven loops.
Twist-2 operators are part of the $\mathfrak{sl}(2)$ sub-sector and contain two scalar fields $\fldZ$ and $\M$ covariant derivatives~$\fldD$
\begin{equation}
\label{twisttwo}
\Tr \left( \fldZ\, \fldD^\M\, \fldZ\,\right)\,.
\end{equation}
There is one primary operator of this type for each even $M$. At one loop at weak coupling, these single-trace operators map to states of the non-compact $\alg{sl}(2)$ spin $=-\sfrac{1}{2}$ length-two Heisenberg magnet with $\M$ excitations. 
The states have the total scaling dimension%
\begin{equation}
\label{dimension}
\Delta=2+\M+\gamma(g)\, ,
\qquad {\rm with} \qquad
\gamma(g)=\sum_{\ell=1}^\infty  \gamma^{}_{2\ell}\,g^{2\ell}\, ,
\end{equation}
where $\gamma(g)$, called the anomalous part of the dimension, depends on the coupling constant
\begin{equation}
\label{convention}
g^2=\frac{\lambda}{16\,\pi^2}\, ,
\end{equation}
and $\lambda=N\, \gym^2$ is the 't Hooft coupling constant. 

From the asymptotic Bethe ansatz \cite{Staudacher:2004tk}, the anomalous dimension $\gamma(g)$ can be determined exactly up to three loops, $\Op(g^6)$.
In the $\mathfrak{sl}(2)$ sector, the asymptotic Bethe equations are \cite{Beisert:2005fw,Beisert:2006ez}
\begin{equation}
\label{sl2eq}
\left(\frac{x^+_k}{x^-_k}\right)^L=\prod_{\substack{j=1\\j \neq k}}^\M
\frac{x_k^--x_j^+}{x_k^+-x_j^-}\,
\frac{1-g^2/x_k^+x_j^-}{1-g^2/x_k^-x_j^+}\,
\exp\big(2\,i\,\theta(u_k,u_j)\big),
\qquad
\prod_{k=1}^M \frac{x^+_k}{x^-_k}=1\, ,
\end{equation}
where the variables $x^{\pm}_k$ are related to the Bethe roots $u_k$ through
\begin{equation}\label{definition x}
x_k^{\pm}=x(u_k^\pm)\, ,
\qquad
u^\pm=u\pm\tfrac{i}{2}\, ,
\qquad
x(u)=\frac{u}{2}\left(1+\sqrt{1-4\,\frac{g^2}{u^2}}\right) \, .
\end{equation}
The anomalous dimension is related to the Bethe roots by
\begin{equation}
\label{dim}
\gaba(g)=2\, g^2\, \sum^{\M}_{k=1}
\left(\frac{i}{x^{+}_k}-\frac{i}{x^{-}_k}\right) =
\sum_{l=1}^\infty g^{2l}\,\gamma^{\ABA}_{2l}(M)\,.
\end{equation}
As we are only interested in the rational part of $\gamma_{14}^{\ABA}(M)$, we put the dressing phase $\theta(u_k,u_j)$ equal to zero, which considerably simplifies our computations.

Our goal is to find a general expression for the anomalous dimension valid at arbitrary $\M$. To do this, we perform perturbative computations at fixed values of $\M$ and match the coefficients to an appropriate ansatz which assumes the maximal transcendentality principle~\cite{Kotikov:2002ab}.
The basis for the ansatz consists of harmonic sums, which can be defined recursively by (see \cite{Vermaseren:1998uu})
\begin{eqnarray} \label{vhs1}
S_a (M)=\sum^{M}_{j=1} \frac{(\mbox{sgn}(a))^{j}}{j^{\vert a\vert}}\, ,\,\,\,\,\,\,\,\,\,\,\,\,\,\,\,\,\,\,\,\,
S_{a_1,\ldots,a_n}(M)=\sum^{M}_{j=1} \frac{(\mbox{sgn}(a_1))^{j}}{j^{\vert a_1\vert}}
\,S_{a_2,\ldots,a_n}(j)\, .\label{vhs}
\end{eqnarray}
To each sum $S_{a_1,\ldots,a_n}$ we assign a \transcendentalitylevel{} $k$, which is given by the sum of the absolute values of its indices
\beq
k=\vert a_1 \vert +\ldots \vert a_n \vert\,,
\eeq
and the \transcendentalitylevel{} of a product of harmonic sums equals the sum of the \transcendentalitylevels{} of its factors.
The maximal transcendentality principle~\cite{Kotikov:2002ab} states that, at a given order of perturbative theory, the anomalous dimension of twist-2 operators contains only harmonic sums with maximal transcendentality. At the $\ell$-loop order, corresponding to \transcendentalitylevel{} $k=2\ell-1$, the dimension of this basis is equal to $((1 - \sqrt{2})^k + (1 + \sqrt{2})^k)/2$, so at seven loops it contains more than 47000 combinations of harmonic sums, see Table~\ref{table:nums}.

\begin{table}[t]
\small
\centering
\begin{tabular}{|c|c|c|c|c|c|c|c|c|c|c|c|c|c}
 \hline
                             &                   &       &       &         &       &          &         &       & & & &     \\[-2mm]
         Contribution        & Rational          & $\z3$ & $\z5$ & $\zp{3}{2}$ & $\z7$ & $\z5\z3$ & $\zp{3}{3}$ & $\z9$ &$\zp{5}{2}$ &$\z7\z3$ &$\zeta_{11}$ & Total\\[2mm]
  \hline
                             &                   &       &       &         &       &          &         &       &  &  &  &      \\[-2mm]
  Transcendentality 
  & 13                &   10   & 8     & 7       & 6     & 5        & 4       & 4     & 3 & 3 & 2 &  \\[2mm]
  \hline
                             &                   &       &       &         &       &          &         &       &  &  &  &      \\[-2mm]
  $\gamma_{14}^{\ABA}$        &  47321             & 3363   & 577    &         & 99    &          &         &    17   &    & & & 51377 \\[1mm]
  $\reciP_{14}^{\ABA}$        &  4096             & 512   & 128    &         & 32     &          &         &    8   &   & & & 4776 \\[2mm]
  $\gamma_{14}^{\rm{\tiny wrap}}$   &  8819             & 577    & 99    & 41      & 17     & 7        & 3       & 3     & 1 & 1& 1& 9569\\[1mm]
  $\reciP_{14}^{\rm{\tiny wrap}}$   &  256              & 32    & 8     & 4       & 2     & 1        & 1       & 1     &   & & & 305\\[2mm]
  \hline
\end{tabular}
\caption{The number of harmonic sums in the basis for contributions of different transcendentality.}
\label{table:nums}
\end{table}

Due to the generalised Gribov-Lipatov reciprocity~\cite{Dokshitzer:2005bf,Dokshitzer:2006nm} the usual harmonic sums (\ref{vhs}) combine into the reciprocity-respecting sums~\cite{Dokshitzer:2006nm,Beccaria:2007bb}, which significantly reduces the dimension of the basis.  The reciprocity-respecting function 
$\reciP(\M)$~\cite{Dokshitzer:2005bf,Dokshitzer:2006nm,Basso:2006nk} is defined by
\begin{equation} \label{Pfunction}
\gamma(M) = \reciP \left(M+\frac{1}{2} \gamma(M) \right)\,
\end{equation}
and is related to the reciprocity-respecting splitting function ${\mathcal P}(x)$~\cite{Dokshitzer:2005bf,Dokshitzer:2006nm} through a Mellin transformation. At at all orders of perturbation theory, ${\mathcal P}(x)$ should satisfy the Gribov-Lipatov relation~\cite{Gribov:1972ri}
\begin{equation}\label{GrLipRel}
{\mathcal P}(x)=-\,x\,{\mathcal P}\!\left(\frac{1}{x}\right)\, .
\end{equation}
An advantage is that $\reciP(M)$ 
can be expressed only in terms of the binomial sums (see \cite{Vermaseren:1998uu})
\beq\label{BinomialSums}
\HBS_{i_1,\ldots,i_k}(N)=(-1)^N\sum_{j=1}^{N}(-1)^j\binom{N}{j}\binom{N+j}{j}\HS_{i_1,...,i_k}(j)\,,
\eeq
and the basis of these sums is equivalent to the basis of the reciprocity-respecting sums\footnote{The relations between the binomial and the nested harmonic sums can be found in the ancillary files of the arXiv version this paper or on the web-page \href{http://thd.pnpi.spb.ru/~velizh/7loop/}{\texttt{http://thd.pnpi.spb.ru/\textasciitilde velizh/7loop/}}.}.
Note that the binomial sums are only defined for positive values of their indices $i_1,\ldots,i_k$. For \transcendentalitylevel{} $k$ the dimension of the basis of binomial sums equals $2^{k-1}$, so at seven loops it contains $2^{13-1}=4096$ binomial harmonic sums (see Table~\ref{table:nums}), which is a significant reduction compared to the basis of harmonic sums in $\gamma_{14}$. 
To compute the rational part of the ABA contribution, we thus need to fix 4096 coefficients, and thus the same number of seven loop solutions of the Bethe equations at fixed $M$ is required. The analytic solution of the Bethe equations for the first 4096 values of $M$ is beyond computer ability, but it was possible to solve these equations up to seven loops numerically with an accuracy of about $10^{-5000}$.
This accuracy is not sufficient to reconstruct the rational numbers that appear in the anomalous dimension for a given $M$, but this is not necessary. What we need is the coefficients that appear in front of the harmonic sums in the general ansatz, and these coefficients were found numerically by using the \texttt{MATHEMATICA} function \texttt{LinearSolve}. The obtained numbers turn out to be very close to integers, and the desired result for $\reciP_{14}(M)$ is given by their rounding. This result can be found in the ancillary files of the arXiv version this paper.
To compute the roots we used the clusters \texttt{HLRN}\footnote{Der Norddeutsche Verbund f\"ur Hoch- und H\"ochstleistungsrechnen \href{https://www.hlrn.de/home/view}{(HLRN)}} and \texttt{CLOU}\footnote{Cluster of UNIX Machines \href{https://www.cms.hu-berlin.de/de/dl/systemservice/computeservice/server/clou/standardseite}{(CLOU)}} and rewrote the initial \texttt{MATHEMATICA} code as a \texttt{GiNaC} code. 

\setcounter{footnote}{7}

Surprisingly, from the analysis of the six-loop anomalous dimension (see Appendix~A in ref.~\cite{Marboe:2014sya}), we have found that some of the binomial sums, which enter in the result for the reciprocity-respecting function, have the same coefficient. This property becomes more clear if we return to the reciprocity-respecting sums instead of the binomial sums. The strong definition of the reciprocity-respecting sums can be found in~\cite{Dokshitzer:2006nm,Beccaria:2007bb}, but as we are only interested in the rational part, we ignore all terms containing $\zeta_i$, which enter into these sums. We denote such sums by $\RRS_{i_1,i_2,\cdots,i_k}$\footnote{The relations between the reciprocity-respecting sums $\RRS_{i_1,i_2,\cdots,i_k}$ 
and the binomial sums can be found in the ancillary files of the arXiv version of this paper.}, where the indices $i_k$ should be only positive odd or negative even numbers (see~\cite{Dokshitzer:2006nm,Beccaria:2007bb}). We have found that the combinations of $\RRS_{i_1,i_2,\cdots,i_k}$, which are related by permutations of the indices $1$ and $3$ inside the subset of indices, which contains only $1$ and $3$, are multiplied by a common factor. For example, the following reciprocity-respecting sums have the same coefficient at six loops:
\begin{eqnarray}
&&\Big(\RRS_{1,1,1,1,3,-4}+\RRS_{1,1,1,3,1,-4}+\RRS_{1,1,3,1,1,-4}+\RRS_{1,3,1,1,1,-4}+\RRS_{3,1,1,1,1,-4}\Big)
\end{eqnarray}
and so on. At seven loops more than 1500 reciprocity-respecting sums will combine into about 200 combinations, which reduces the basis to less than 2700 terms. We therefore tried to find coefficients in front of terms in the redefined basis numerically, i.e. solving a system of about 2700 equations on 2700 variables, and we found that this system has a unique solution with coefficients that are numerically very close to integer numbers.

The seven-loop contribution to the reciprocity-respecting function
\begin{equation}
\reciP^{\ABA}(M)=\sum_{l=1}^\infty g^{2l}\,\reciP^{\ABA}_{2l}(M)\,.\label{PABA}
\end{equation}
is related to the anomalous dimension in the following way:
\beqa
&&\hspace*{-12mm}\hat\reciP_{14}\ =\ 
\hat\gamma_{14}
-\frac{1}{2} \left(\hat\gamma_6\hat\gamma_8
+\, \hat\gamma_4 \hat\gamma_{10}
+\, \hat\gamma_2 \hat\gamma_{12}\right)'
+\frac{1}{8} \left(
\hat\gamma_4^2\hat\gamma_6
+\hat\gamma_2\hat\gamma_6^2
+2\, \hat\gamma_2\hat\gamma_4\hat\gamma_8
+\hat\gamma_2^2 \hat\gamma_{10}\right)''\nonumber\\
&& \hspace*{-10mm}
-\frac{1}{24} \left(
 \hat\gamma_2\hat\gamma_4^3
+3\, \hat\gamma_2^2 \hat\gamma_4\hat\gamma_6
+\, \hat\gamma_2^3 \hat\gamma_8
\right)'''
+\frac{5}{384} \left(
2\,  \hat\gamma_2^3\hat\gamma_4^2
+\, \hat\gamma_2^4 \hat\gamma_6\right)''''
-\frac{\left(\hat\gamma_2^5 \hat\gamma_4\right)'''''}{3840}
+\frac{\left(\hat\gamma _2^7\right)''''''}{322560}\,,
\label{P14}
\eeqa
where $\hat{\reciP}_{2\ell}={\reciP}_{2\ell}^{\ABA,\,\rm rational}(\M)$, $\hat{\gamma}_{2\ell}={\gamma}_{2\ell}^{\ABA,\,\rm rational}(\M)$ and each prime marks a derivative with respect to $\M$.

The final expression for the rational part of the ABA contribution to the seven-loop anomalous dimension of twist-2 operators in the canonical basis of the usual harmonic sums~(\ref{vhs}) and the result for ${\reciP}_{14}^{\ABA,\,\rm rational}(\M)$ can be found in the ancillary files.

\section{Full seven-loop anomalous dimension at fixed $M$} \label{sec:finiteS}

The quantum spectral curve \cite{Gromov:2013pga,Gromov:2014caa} is currently the most concise integrability-based formulation of the all-loop spectral problem of the AdS$_5$/CFT$_4$ correspondence.
For twist-2 operators, the applications of the QSC has led to new results in the small spin limit \cite{Gromov:2014bva}, in the BFKL regime \cite{Alfimov:2014bwa,Gromov:2015vua}, at weak coupling \cite{Marboe:2014gma}, and numerically at any coupling \cite{Gromov:2015wca}.
The QSC formulates the spectral problem in terms of a $Q$-system, which is a fundamental structure in integrable models.
The involved $Q$-functions all depend on the spectral parameter and are related by finite difference equations. Furthermore, the QSC specifies the analytic structure and asymptotic behavior at large spectral parameter of these functions. Each superconformal multiplet of single-trace operators in $\mathcal{N}=4$ SYM correspond to a particular solution of this system with certain boundary conditions in the form of large $u$ asymptotics dictated by the weights of the operators with respect to the superconformal algebra.

The perturbative solution of the system involves solving a number of finite difference equations at each order. The involved operations introduce only a very limited set of functions: rational functions and so-called $\eta$-functions of the spectral parameter.
The procedure is initialized by the solution of Bethe/Baxter equations (or equivalent) and this is the only point where irrational algebraic numbers are, possibly, introduced.

For general operators, the known perturbative solution methods of the QSC involves the full $psu(2,2|4)$ $Q$-system. However, for operators belonging to the $sl(2)$ sector, it is convenient to consider only a subset of the QSC, the so-called $\mathbf{P}\mu$-system, a closed set of 9 independent functions \cite{Marboe:2014gma}.

\subsection*{Solving the $\bP\mu$-system perturbatively}
The $\bP\mu$-system contains a $4\times 4$ antisymmetric matrix $\mu_{ab}$ and four functions $\bP_a$ satisfying the equations 
\begin{eqnarray}
\mu_{ab}-\tilde\mu_{ab}= \tilde \bP_a\bP_b-\tilde \bP_b\bP_a\,,\quad\quad
\tilde \bP_a = \mu_{ab}\chi^{bc}\,\bP_c\,,\quad\quad
\tilde\mu_{ab}(u) =\mu_{ab}(u+i)\,, \label{pmueq}
\end{eqnarray}
where $\chi^{ab}=\text{antidiag}(-1,1,-1,1)$ and a tilde denotes the analytic continuation through a branch cut on the real axis.
The functions are multi-valued functions of the spectral parameter, $u$, and have branch points at $\pm 2g + i \mathbb{Z}$. With short cuts between these branch points, the functions $\bP_a$ have branch points only at $u=\pm 2g$ on their first Riemann sheet.
For $g\to 0$ the branch points collide into points on the imaginary axis, and these are the only points where the functions are allowed to be singular. Combined with their powerlike asymptotics at $u\to\infty$, this means that $\bP_a$ are rational functions of $u$ to all orders in perturbation theory.

From \eqref{pmueq} it is possible to derive a second order difference equation on $\mu_{12}$. For $sl(2)$ operators this equation is homogeneous at the leading order due to the important property $\bP_1=\mathcal{O}(g^2)$ and reads
\begin{eqnarray}\label{BaxterFull}
\frac{1}{\bP_2^2}\,\mu_{12}-\left(\frac{\bP_3}{\bP_2}-\frac{\bP_3^{[2]}}{\bP_2^{[2]}}+\frac 1{\bP_2^2}+\frac 1{\left(\bP_2^{[2]}\right)^2}\right)\mu_{12}^{[2]}+\frac 1{\left(\bP_2^{[2]}\right)^2}\,\mu_{12}^{[4]}
&=& 0 \,, 
\end{eqnarray}
where we used the notation $f^{[n]}(u)\equiv f(u+\frac{in}{2})$. For twist-2 operators, the $u\to\infty$ asymptotics furthermore sets $\bP_2=\frac{A_2}{u}+\mathcal{O}(g^2)$, $\bP_3=A_3+\mathcal{O}(g^2)$ and $A_2A_3=-iM(M+1) + \mathcal{O}(g^2)$. If we identify the leading contribution to $\mu_{12}$ with the Baxter polynomial through $\mu_{12}(u)\propto Q(u-\frac{i}{2})$, this is exactly the well-known 1-loop Baxter equation. 

One of the key steps in the algorithm is to solve an inhomogeneous version of this equation at each loop, and in that sense, the QSC approach is somewhat similar in philosophy to the ABA approach. The difference is that the perturbative corrections to $\mu_{12}$ are not rational functions of $u$.

The conformal dimension, along with the remaining quantum numbers, enters through the asymptotics of $\bP_a$ and is fixed order by order as part of the algorithm. For example, the seven-loop anomalous dimension for $M=4$ is found to be
\begin{eqnarray}
\gamma_{14}(4)&=&
-\frac{25166596925125}{4251528}
+\frac{290741688625}{19683}\zeta_3
+\frac{11516727625}{4374}\zeta_5
-\frac{1808233750}{729}\zeta_3^2 \nonumber\\
&& -\frac{17907365875}{2916}\zeta_7
+\frac{975687500}{243}\zeta_3\zeta_5
-\frac{1756580750}{243}\zeta_9
+\frac{12500000}{27}\zeta_3^3 \nonumber\\
&&-\frac{71875000}{27}\zeta_5^2
-\frac{140000000}{27}\zeta_3\zeta_7
+\frac{42350000}{3}\zeta_{11}\,.
\end{eqnarray}

\subsection*{Working with partial results}
With increasing $M$, the complexity of the functions $\mu_{ab}$ grows. For example, the function $\mu_{12}$ is a polynomial of order $M$ at the leading order. This means that the computation time and memory usage also grows significantly with $M$, and it limits the results that are within reach, even on high-performance computer clusters. To be able to generate enough data, we exploited the fact that it is possible to work with only partial results.

For fixed $M$, all functions are built from parts that are proportional to different $\zeta$-values, e.g.
\begin{eqnarray}
\bP_a= \bP_a^{\text{rational}} + \zeta_3 \bP_a^{\zeta_3} + \zeta_5 \bP_a^{\zeta_5}+\zeta_3^2 \bP_a^{\zeta_3^2}+\hdots
\end{eqnarray}
All operations in the algorithm simply multiply these terms, so a term proportional to $\zeta_3$ will never contribute to the part without $\zeta$-value dependence, a term proportional to $\zeta_5$ will never influence the $\zeta_3$ part, etc.

This means that it is possible to run the algorithm keeping only parts of the results, and the obtained partial functions still satisfy the analytical requirements imposed in the algorithm. To generate more results for the reconstruction of $\gamma_{14}^{\text{rational}}(M)$ and $\gamma_{14}^{\zeta_3}(M)$, we have used this property in two modifications of our \texttt{Mathematica}-implementation of the algorithm: one keeping only the $\zeta$-value independent part of the results, and another keeping also the part proportional to $\zeta_3$. Sample computation times for the different versions of the code are given in Table \ref{table:time}.

We computed the full seven-loop anomalous dimension for the 32 lowest even integer spins, i.e. for $M=2,4,\hdots,64$. This is enough to reconstruct all $\zeta_i$ contributions to $\cP_{14}$ except for the $\z3$ and rational contributions. To be able to reconstruct $\cP_{14}^{\z3}$, we additionally computed the rational and $\z3$ contributions for the next 25 even integer values ($M=66,68,\hdots,114$). Finally, the rational contribution for all even integers up to $M=290$ were calculated in order to be able to reconstruct the wrapping contribution to $\cP_{14}^{\text{rational}}$.

\begin{table}[t]
\centering
\begin{tabular}{|c|
c|c|c|c|c|c|c|c|c|c|c|}
\hline
$M$&2&4&6&8&10&20&30&40&50&60&70\\\hline
full&1&2.1&3.3&4.8&7.0&27&113&&&&\\
rational and $\zeta_3$ &0.55& 1.3& 1.8& 2.7& 3.7& 11 & 26 & 51 &87&132 & 201\\
rational&0.33&0.65&1.0&1.5&2.0&6.4&14&26&45&74&108\\\hline
\end{tabular}
\caption{\small Computation time normalized by the $M=2$ computation time for the full result (120 seconds on a standard laptop). Note that the reduced codes use significantly less memory than the full code which makes it possible to go to much higher $M$ before lack of memory becomes an issue.}
\label{table:time}
\end{table}

\section{Reconstruction of the seven-loop anomalous dimension at arbitrary $M$} \label{sec:L7}

As described in Section~\ref{sec:ABA} we will reconstruct the reciprocity function $\reciP(\M)$, which has the following structure at seven loops
\beqa
\cP_{14}&=&
\cP_{14}^{{\mathrm{rational}}}
+\z3\cP_{14}^{\z3}
+\z5\cP_{14}^{\z5}
+\zp{3}{2}\cP_{14}^{\zp{3}{2}}
+\z7\cP_{14}^{\z7}
+\z3\z5\cP_{14}^{\z3\z5}
+\zp{3}{3}\cP_{14}^{\zp{3}{3}}\nonumber\\
&&\hspace*{10mm}
+\z9\cP_{14}^{\z9}
+\zp{5}{2}\cP_{14}^{\zp{5}{2}}
+\z3\z7\cP_{14}^{\z3\z7}
+\zeta_{11}\cP_{14}^{\zeta_{11}}
\label{P14AnsatzZs}
\eeqa
and is related to $\gamma_{14}$ through eq.~(\ref{P14}) with the replacement of $\hat{\gamma}_{2\ell}$ by the full $\ell$-loop anomalous dimension $\gamma_{2\ell}$.

The basis for $\cP_{14}$ consists of the binomial harmonic sums~(\ref{BinomialSums}) and the number of such sums in the corresponding basis are listed in the Table~\ref{table:nums}. Using the first 32 even values of the full seven-loop anomalous dimension we found the results for all $\cP_{14}^{\zeta_i}$ down to $\cP_{14}^{\z7}$ exactly, while $\cP_{14}^{\zp{3}{2}}$ and $\cP_{14}^{\z5}$ was reconstructed with the help of the LLL-algorithm. $\cP_{14}^{\z3}$, which has $2^{10-1}=512$ binomial harmonic sums in the basis, was reconstructed with the help of the LLL-algorithm from the first 57 even values. The simplest $\cP_{14}^{\zeta_i}$ have the following form:
\beqa
\cP_{14}^{\zeta_{11}}&=&813120\,\HBS_1^2\,,\label{Resultz11}\\
\cP_{14}^{\zp{5}{2}}&=&-36800\,\HBS_1^3\,,\label{Resultz52}\\
\cP_{14}^{\z7\z3}&=&-71680\,\HBS_1^3\,,\label{Resultz7z3}\\
\cP_{14}^{\zp{3}{3}}&=&1536\,\HBS_1^4\,,\label{Resultz33}\\
\cP_{14}^{\z9}&=&
-\frac{64}{3}\, \HBS_1 
\left(
-11424\, \HBS_{2,1}
+1901\, \HBS_1^3
+24444\, \HBS_1 \HBS_2
-10332\, \HBS_3
\right)
\,,\label{Resultz9}\\
\cP_{14}^{\z5\z3}&=&
\frac{64}{3}\, \HBS_1^2 
\left(
-892\, \HBS_{2,1}
+253\, \HBS_1^3
+3426\, \HBS_1 \HBS_2
-2532\, \HBS_3
\right)
\label{Resultz5z3}
\eeqa
and all other contributions to $\cP_{14}$ can be found in the ancillary files of the arXiv version of the paper. 

For the reconstruction of the rational part, we propose the following general ansatz for the basis of $\cP_{14}^{\Wrap}$
\beqa
{\mathrm{Basis}}\Big[\cP_{14}^{\Wrap}\Big]&=&
\Big\{
\cP_2^2\, \cT_{14},
\Big[\cP_2 \cP_4 \cT_{12}\Big],
\Big[\cP_2^4 \cT_{12}\Big],
\Big[\cP_2 \cP_6\cT_{10}\Big],
\Big[\cP_2^3 \cP_4\cT_{10}\Big],
\Big[\cP_2^6\cT_{10}\Big],
\Big[\cP_4^2\cT_{10}\Big],
\nonumber\\
&&
\Big[\cP_2 \cP_8\cT_{8}\Big],
\Big[\cP_2\cP_4^2\cT_{8}\Big],
\Big[\cP_2^3 \cP_6\cT_{8}\Big],
\Big[\cP_2^5 \cP_4\cT_{8}\Big],
\Big[\cP_2^8\cT_{8}\Big],
\Big[\cP_4 \cP_6\cT_{8}\Big]
\Big\}
\,,
\ \qquad\label{P14Ansatz}
\eeqa
where $\cP_{2\ell}$ is the $\ell$-loop reciprocity function of the full $\ell$-loop anomalous dimension $\gamma_{2\ell}$, and
$\cT_{2\ell}$ is the part of the $\ell$-loop $\cP_{2\ell}^{\Wrap}$, which contains the products of binomial harmonic sums including $\cP_2^k\sim\HBS_1^k$ with $k\geq2$, for example, $ \cT_{8}$ is defined as:
\beqa
&&\gamma^{\Wrap}_{8}=\reciP_{8}^{\Wrap}=\reciP_{2}^{2}\,\cT_{8}\,,\\
&&\cT_{8}=\Big(-5\,\z5 +2\,\HBS_{2}\,\z3+\left(\HBS_{2,1,2}-\HBS_{3,1,1}\right)\Big)\,,\\
&&\gamma_{2}=\cP_{2}=4\,\HBS_1\,.
\eeqa
and $\cT_{10}$ and $\cT_{12}$ can be found in ref.~\cite{Lukowski:2009ce} and ref.~\cite{Marboe:2014sya} correspondingly. 
The square brackets on the right hand side of the equation mean that all terms in the expansion of the expression will enter in the basis, e.g.
\beqa
\Big[\cP_2\cP_4\cT_{8}\Big]&=&\Big[\HBS_1\big(\HBS_1\HBS_2-\HBS_3-\HBS_{2,1}\big)\big(\HBS_{2,1,2}-\HBS_{3,1,1}\big)\Big]=\nonumber\\
&&\hspace*{-13mm}=
\Big\{
\HBS_1\HBS_1\HBS_2\HBS_{2,1,2},
\HBS_1\HBS_3\HBS_{2,1,2},
\HBS_1\HBS_{2,1}\HBS_{2,1,2},
\HBS_1\HBS_1\HBS_2\HBS_{3,1,1},
\HBS_1\HBS_3\HBS_{3,1,1},
\HBS_1\HBS_{2,1}\HBS_{3,1,1}
\Big\}.\qquad\label{ToBasis}
\eeqa
The function $\cT_{14}$ from eq.~(\ref{P14Ansatz}) has the transcedentality $11$ and the basis for this structure consists of $2^{11-1}=1024$ binomial harmonic sums. Some terms in eq.~(\ref{P14Ansatz}) are linearly dependent and in total they add about 400 binomial harmonic sums, i.e. our basis for $\cP_{14}^{\Wrap}$ contains about 1400 binomial harmonic sums.

With the QSC method we computed the anomalous dimension $\gamma(M)$ for the first 145 even values: $M=2,\ 4,\ \cdots,\ 290$. The direct application of the LLL-algorithm with the \texttt{fplll}-code did not give us any reasonable result, spending about a month for the LLL-reduction with the fastest set of parameters\footnote{The LLL-algorithm and its \texttt{fplll} realisation have some parameters, which can be changed to choose accuracy/time for the reduction procedure. The default value for the $\eta$-parameter in the \texttt{fplll}-code is equal to $0.51$ (see manual for \texttt{fplll}-code~\cite{fplll}), the same as in the original LLL-algorithm, while we have found, by trying on lower loops, that for our purpose it is possible to set this parameter to $0.98$, which increases the speed of the LLL-reduction drastically.}. To speed up the computation, we would either need to produce more data points for $\gamma(M)$ or optimise the \texttt{fplll}-code, which can in both cases be done by parallelisation. However, we found a simpler way to find the desired results without such complications.

The idea that we used is based on the fact that the general expression for the anomalous dimension of twist-2 operators with arbitrary $M$ can be rewritten in the following form:
\beqa
\gamma(M)&=&\sum_{i}g^{2i}\,\gamma_{2i}(M)=\sum_{i}g^{2i}\sum_{\veci=2i-1} \cC^j_{i_1,\ldots,i_k}\ S_{i_1,\ldots,i_k}\nonumber\\
&=&\sum_{i}g^{2i}\sum_{j=0}^{2i-1}S_1^j\sum_{\substack{\veci=2i-1-j\\ i_1\neq 1}} \cC^j_{i_1,\ldots,i_k}\ S_{i_1,\ldots,i_k}\,.
\eeqa
In other words, we can extract the powers of the simplest harmonic sum $S_1$ multiplied by sums for which the first index is not equal to $1$. In our case, for $\cP_{14}^{\Wrap}$ we have 
\beqa
\cP_{14}^{\Wrap}
&=&\sum_{j=0}^{13}\HBS_1^j\sum_{\substack{\veci=13-j\\ i_1\neq 1}} \cC^j_{i_1,\cdots,i_k}\ \HBS_{i_1,\cdots,i_k}\,.
\label{P14cC}
\eeqa
So, our initial task is subdivided into several parts, and we will reconstruct the parts with different powers of $\HBS_1$ separately. If we are able to find the part which does not contain $\HBS_1$ at all, we can subtract it from the full result and factorize $\HBS_1$ in the remaining part:
\beqa
\cP_{14}^{\Wrap}-\sum_{\substack{\veci=13\\ i_1\neq 1}} \cC^0_{i_1,\cdots,i_k}\ \HBS_{i_1,\cdots,i_k}
&=&\HBS_1\sum_{j=0}^{12}\HBS_1^j\sum_{\substack{\veci=12-j\\ i_1\neq 1}} \cC^j_{i_1,\cdots,i_k}\ \HBS_{i_1,\cdots,i_k}\,.\label{LeftS1}
\eeqa
This means that the left-hand side of eq. (\ref{LeftS1}), in which $\cP_{14}^{\Wrap}$ is known for fixed values of $M$, should be divisible by $\HBS_1$ with the same $M$, and as a consequence the numerator of this rational number should be divisible by the numerator of $\HBS_1(M)$.
In this way we put on the left-hand side of eq. (\ref{LeftS1}) all binomial harmonic sums from the basis (\ref{P14Ansatz}) without $\HBS_1$ with unknown coefficients $\cC^0_{i_1,\cdots,i_k}$, compute the result for each $M$, and find the common denominator. Then take the numerator from the obtained expression and divide it by the numerator of the corresponding $\HBS_1(M)$ - the obtained number should be an (small) integer number, when we substitute all unknown coefficients $\cC^0_{i_1,\cdots,i_k}$. As the numbers in the front of the coefficients $\cC^0_{i_1,\cdots,i_k}$ are huge numbers (usually prime numbers), we can work with the remainders of the initial numbers modulo the numerator of the corresponding $\HBS_1(M)$
\beqa
&&\bigg(\left[\cP_{14}^{\Wrap}\right]\ \mathrm{mod}\ {\mathfrak{N}[\HBS_1]}\bigg)
-\sum_{\substack{\veci=13\\ i_1\neq 1}} \cC^{\,0}_{i_1,\cdots,i_k}\ 
\bigg(\left[\HBS_{i_1,\cdots,i_k}\right]\ \mathrm{mod}\ {\mathfrak{N}[\HBS_1]}\bigg)= c^0_M\,{\mathfrak{N}[\HBS_1]}
\,.\label{LeftS1mod}
\eeqa
where $\mathfrak{N}[\HBS_1]$ means the numerator of $\HBS_1$ and values for $\cP_{14}^{\Wrap}$, $\HBS_{i_1,\cdots,i_k}$ and $\HBS_1$ are taken for the same $M$. For each $M$,
$c_M^0$ is an integer number to be fixed along with $\cC^0_{i_1,\cdots,i_k}$.
$\left[\cP_{14}^{\Wrap}\right]$ and $ \left[\HBS_{i_1,\cdots,i_k}\right]$ denote the values of $\cP_{14}^{\Wrap}$ and $\HBS_{i_1,\cdots,i_k}$ when their common denominator is factored out, that is in \texttt{MATHEMATICA} language we apply the function \texttt{Together} to the expression in the square brackets, factor out the common denominator, and use the expression in the round brackets:
\beqa
&&\mathtt{Together}\left[\cP_{14}^{\Wrap}
-\sum_{\substack{\veci=13\\ i_1\neq 1}} \cC^{\,0}_{i_1,\cdots,i_k}\ 
\HBS_{i_1,\cdots,i_k}\right]=\nonumber\\
&&\frac{1}{\mathtt{[Denominator]}}\left(\left[\cP_{14}^{\Wrap}\right]
-\sum_{\substack{\veci=13\\ i_1\neq 1}} \cC^{\,0}_{i_1,\cdots,i_k}\ 
\left[\HBS_{i_1,\cdots,i_k}\right]\right)
\,.\label{eq:Together}
\eeqa

We evaluated the expression (\ref{LeftS1mod}) for $M=2,\ 4,\ \cdots,\ 290$ and applied the LLL-algorithm with the \texttt{fplll}-code for the obtained system of linear Diophantine equations on $62$ coefficients $\cC^{\,0}_{i_1,\cdots,i_k}$ (the number of the binomial harmonic sums in the basis from eq. (\ref{P14Ansatz})) and $145$ coefficients $c^0_M$, i.e. a system of $145$ linear equations on 207 unknowns. The obtained LLL-reduced matrix contains a line with the desired coefficients $\cC^{\,0}_{i_1,\cdots,i_k}$.

In the next step we look for the coefficients $\cC^{\,1}_{i_1,\cdots,i_k}$, which are proportional to $\HBS_1$
\beqa
\cP_{14}^{\Wrap}
&-&\sum_{\substack{\veci=13\\ i_1\neq 1}} \cC^{\,0}_{i_1,\cdots,i_k}\ \HBS_{i_1,\cdots,i_k}
\ -\ \HBS_1\sum_{\substack{\veci=12\\ i_1\neq 1}} \cC^{\,1}_{i_1,\cdots,i_k}\ \HBS_{i_1,\cdots,i_k}\ =\ \nonumber\\
&&\hspace*{20mm}\ =\ \HBS_1^2\sum_{j=0}^{11}\HBS_1^j\sum_{\substack{\veci=11-j\\ i_1\neq 1}} \cC^j_{i_1,\cdots,i_k}\ \HBS_{i_1,\cdots,i_k}\,.\label{LeftS1p2}
\eeqa
For this case eq. (\ref{P14Ansatz}) gives us a basis with $327$ combinations of binomial harmonic sums. However, from the analysis of the corresponding result for the $\zeta_3$-contribution, we have found that eq. (\ref{P14Ansatz}) does not give all necessary binomial harmonic sums in the basis and the missing sums can be generated from the following outer product
\beqa
\HBS_1\otimes\Big[\cT_{10}\Big]\otimes \Big\{\HBS_{5},\ \HBS_{3,2},\ \HBS_{2,3},\ \HBS_{2,1,2},\ \HBS_{3,1,1}\Big\}\,,
\eeqa
where $\Big[\cT_{10}\Big]$ denotes all combinations of binomial harmonic sums, which are contained in $\cT_{10}$ (see eq. (\ref{ToBasis})). This gives $14$ additional terms in the basis (other terms will be linearly dependent). To find the coefficients $\cC^{\,1}_{i_1,\cdots,i_k}$ we have $145$ equations on $341+145$ unknowns and the \texttt{fplll}-code produces the LLL-reduced matrix after about $20$ hours on a standard computer.

Proceeding in the same way for all other terms, which are proportional to $\HBS_1^2,\ \HBS_1^3\, \ldots$ we are able to find all coefficients $\cC^{j}_{i_1,\cdots,i_k}$. Finally, we check that eq. (\ref{P14cC}) is indeed satisfied.

\section{Weak coupling constraints}\label{sec:weak}

Having reconstructed the full seven-loop anomalous dimension at arbitrary $M$, we can check the consistency of our result by analytical continuation to negative spin, where known constraints apply. We will consider three classes of constraints coming from the BFKL equation and from the generalised double-logarithmic equation at $\M=-2+\omega$ and at $\M=-r+\omega$, where $r=4,\,6,\,8,...$. The analytic continuation can be done with the help of  \texttt{HARMPOL}~\cite{Remiddi:1999ew} and \texttt{SUMMER} packages~\cite{Vermaseren:1998uu} for \texttt{FORM}~\cite{Vermaseren:2000nd}.
At one loop, the analytic continuation is straightforward since
\beq
\gamma_{2}(M) = 8\,g^2\,S_1 (M) = 8\,g^2\, \left(\Psi(M+1)-\Psi(1)\right)\,,
\eeq
where $\Psi(x)=\frac{d}{dx}\,\log \Gamma(x)$ is the digamma function.
At any loop order singularities are expected to appear at all negative integer values of $\M$.

\subsection{BFKL equation} \label{sec:BFKL}

\setcounter{footnote}{0}

The first in this series of singular points,
\begin{equation}
\label{omega}
M=-1+\omega\,,
\end{equation}
where $\omega$ is infinitesimal,
corresponds to the so-called Balitsky-Fadin-Kuraev-Lipatov (BFKL) pomeron. The BFKL equation \cite{Lipatov:1976zz,Kuraev:1977fs,Balitsky:1978ic} relates $\gamma(g)$ and $\omega$, and it predicts that, when expanded in $g$, the $\ell$-loop anomalous dimension $\gamma_{2 \ell} (\omega)$ exhibits poles in $\omega$. The residues and the order of these poles can be derived directly from the BFKL equation. The BFKL equation has been formulated up to the next-to-leading logarithm approximation (NLLA) \cite{Fadin:1998py,Kotikov:2000pm} and next-to-next-to-leading logarithm approximation (NNLLA) \cite{Gromov:2015vua,Velizhanin:2015xsa,Caron-Huot:2016tzz} and determines the leading, next-to-leading and next-to-next-to-leading poles of $\gamma_{2\ell}(\omega)$. So, we can control the three highest poles of the analytically continued anomalous dimension at $\M=-1+\omega$, which have the following form:
\begin{eqnarray}
 \gamma&=&
\left(2+0\,\omega-2\,\z2\,\omega^2
\right)
\left(\frac{-4\,g^2}{\omega}\right) 
+\left(0+0\,\omega
+4\,\z3\,\omega^2
\right)\,\left(\frac{-4\,g^2}{\omega}\right)^2
\nonumber
\\[-0.2mm]
&&
+\left(0+\,\z3\,\omega-\frac{29}{4}\,\z4\,\omega^2
\right)\,\left(\frac{-4\,g^2}{\omega}\right)^3\nonumber \\[-0.2mm]
&&
+\left(-4\,\z3-\frac{5}{4}\,\z4\,\omega+\left(5\,\z2\z3+\frac{77}{4}\,\z5\right)\,\omega^2 \right)\,\left(\frac{-4\,g^2}{\omega}\right)^4 \nonumber \\[-0.2mm]
&&
+\left(0-\bigg(2\,\z2\,\z3+16\,\z5\bigg)\,\omega
-\left(21\,\zp{3}{2}\,\z3+\frac{61}{3}\,\z6\right)\,\omega^2
\right)\left(\frac{-4g^2}{\omega}\right)^5 \nonumber\\[-0.2mm]
&&
+\left(
-
4\,\z5
-\left(3\,\zp{3}{2}-\frac{143}{48}\,\z6\right)\,\omega
+\left(\frac{277}{3}\,\z3\,\z4
  +8\,\z2\,\z5
  -\frac{631}{32}\,\z7\right)\,\omega^2
\right)\left(\frac{-4g^2}{\omega}\right)^6\nonumber\\[-0.2mm]
&&
+\left(
24\,\zp{3}{2}
-\left(\frac{25}{2}\z3\z4 -2 \z2\z5-38\z7\right)\,\omega\right.\nonumber\\[-0.2mm]
&&\hspace*{3mm}\left.
+\left(
 \frac{32}{171}\,\h_{5,3}
 -\frac{160}{57}\,\h_{7,1}
 -51\,\z2\,\zp{3}{2}
 -\frac{8127}{38}\,\z3\,\z5
 +\frac{91543}{5472}\,\z8
\right)\,\omega^2
\right)\left(\frac{-4g^2}{\omega}\right)^7,
\label{BFKLPredictions}
\end{eqnarray}
where
\beq
\h_{i_1,i_2,...,i_k}=H_{-i_1,-i_2,...,-i_k}(1)
\eeq
and $H_{i_1,...,i_k}(x)$ are the harmonic polylogarithms~\cite{Remiddi:1999ew}.
The last line of the above equation exactly matches the prediction written in the paper~\cite{Velizhanin:2015xsa}\footnote{S. Caron-Huot informed us that the result for the expansion of the NNLLA eigenvalues of the kernel of the BFKL equation, which is given explicitly in \cite{Velizhanin:2015xsa}, is the same as for the results presented in refs.~\cite{Gromov:2015vua} and \cite{Caron-Huot:2016tzz}.}.

\subsection{Generalised double-logarithmic equation at $\M=-2+\omega$} \label{sec:GenerDL2}

Further constraints on the anomalous dimension arises in the vicinity of $M=-2$. These constraints are related to the double-logarithmic asymptotics of scattering amplitudes, which were studied in QED and QCD in \cite{Gorshkov:1966qd,Kirschner:1983di} 
(see also the {\texttt{arXiv}} version of \cite{Kotikov:2002ab}). 
The double-logarithmic equation has the form
\begin{equation}\label{DL}
\gamma\,(2\,\omega+\gamma)=-16\,g^2\,,
\end{equation}
and its solution predicts the highest pole $(g^{2k}/\omega^{2k-1})$ in all orders of perturbation theory:
\begin{eqnarray}\label{dlevenp}
\gamma&=&-\omega+\omega\, \sqrt{1-\frac{16 g^2}{\omega^2}}
=
2\,\frac{(-4\, g^2)}{\omega}
-2\,\frac{(-4\, g^2)^2}{\omega^3}
+4\,\frac{(-4\, g^2)^3}{\omega^5}
-10\,\frac{(-4\, g^2)^4}{\omega^7}\nonumber \\
&&\hspace*{40mm}
+28\,\frac{(-4\, g^2)^5}{\omega^9}
-84\,\frac{(-4\, g^2)^6}{\omega^{11}}
+264\, \frac{(-4\, g^2)^7}{\omega^{13}}
+\ldots\, .\label{DLSolve}
\end{eqnarray}

Motivated by the study of the analytic properties of the anomalous dimension of twist-2 operators in $\cN=4$ SYM theory, a simple generalisation of the double-logarithmic equation has been suggested~\cite{Velizhanin:2011pb}\footnote{Originally, a such generalisation was suggested by L. N. Lipatov and A. Onishchenko in 2004,
but was not published. It was later improved by L. N. Lipatov in ref.~\cite{Kotikov:2007cy}.}.
The proposal is that only the right-hand side of the leading order equation~(\ref{DL}) is modified, and that this modification, besides an expansion in the coupling constant $g^2$, consists of only regular terms depending on $\omega$ (and, possibly, $\gamma$).

By substituting the analytic continuation to the regime $\M=-2+\omega$ of our result for the anomalous dimension into eq.~(\ref{DL}) we indeed obtain the following form of the generalised double-logarithmic equation~\cite{Velizhanin:2011pb}\footnote{We used \texttt{DATAMINE}~\cite{Blumlein:2009cf} tables for the substitution of the multiple zeta functions, or multiple polylogarithms at $x=1$ through usual Euler zeta-functions $\zeta_i$ and the minimal numbers of multiple zeta-functions up to weights $12$.}
{\allowdisplaybreaks
\begin{eqnarray}\label{DLgener}
\gamma\,(2\,\omega+\gamma)&=&
-16\, g^2
-64\, g^4 \z2
+g^6 (128\, {\z3}+256\, {\z4})\nonumber\\
&+&g^8 \left(2560\, {\z2}\, {\z3}+384\, {\zp{3}{2}}-128\, {\z5}+\frac{1888}{3}\, {\z6}\right)
\nonumber\\
&+&g^{10} \left(
-\frac{114688}{171}\,{\h_{5,3}}
+\frac{573440}{57}\,{\h_{7,1}}
+7168\,{\z2}\,{\zp{3}{2}}
-37888\,{\z2}\,{\z5}\right.\nonumber\\
&&\left.
+19456\,{\z3}\,{\z4}
-\frac{237568}{19}\,{\z3}\,{\z5}
-11520\,{\z7}
-\frac{2198944 }{171}\,{\z8}
\right)\nonumber\\
&+&g^{12} 
\Bigg(
\frac{26968064 }{4749}\,{\h_{7,3}}
-\frac{755105792}{4749} \,{\h_{9,1}}
-\frac{3276800}{513} \,{\h_{5,3}}\,{\z2}
+\frac{16384000}{171} \,{\h_{7,1}} \,{\z2}
\nonumber\\
&&
+\frac{44565768064 }{270693}\,{\zeta_{10}}
-12288 \,{\z2} \,{\zp{3}{2}}
-\frac{13583360}{57} \,{\z2} \,{\z3} \,{\z5}
+486400 \,{\z2} \,{\z7}\nonumber\\
&&
-13312 \,{\zp{3}{3}}
+96768  \,{\z4}\,{\zp{3}{2}}
-4096 \,{\z3} \,{\z5}
-\frac{384512}{3} \,{\z3} \,{\z6}
+\frac{254253056}{1583} \,{\z3} \,{\z7}\nonumber\\
&&
-464384 \,{\z4}\,{\z5}
+\frac{123298944}{1583} \,{\zp{5}{2}}
+\frac{2388992 }{9}\,{\z9}\Bigg)\nonumber\\
&+&g^{14} \Bigg(
-\frac{5523846987776}{223994619} \,{\h_{5,3}} \,{\z3}
-\frac{32178176}{513} \,{\h_{5,3}}\,{\z4}
-\frac{1150303600640 }{74664873}\,{\h_{5,3,3}}\nonumber\\
&&
-\frac{572512010240 }{8296097}\,{\h_{5,5,1}}
+\frac{30965188526080}{74664873} \,{\h_{7,1}} \,{\z3}
+\frac{160890880}{171} \,{\h_{7,1}} \,{\z4}\nonumber\\
&&
+\frac{1421475840000}{8296097} \,{\h_{7,1,3}}
+\frac{116916224}{1583} \,{\h_{7,3}} \,{\z2}
+\frac{1917059072000}{8296097} \,{\h_{7,3,1}}\nonumber\\
&&
-\frac{210501632}{47475} \,{\h_{7,5}}
-\frac{3273654272}{1583} \,{\h_{9,1}} \,{\z2}
-\frac{13601813299200 }{8296097}\,{\h_{9,1,1}}\nonumber\\
&&
-\frac{7340032}{211} \,{\h_{9,3}}
+\frac{7901544448}{3165} \,{\h_{11,1}}
+\frac{59192446673984 }{24888291}\,{\zeta_{11}}\nonumber\\
&&
-\frac{443833733896402112 }{986682752325}\,{\zeta_{12}}
-471040 \,{\z2} \,{\zp{3}{3}}
+286720 \,{\z2} \,{\z3} \,{\z5}\nonumber\\
&&
+\frac{5494323712 }{1583}\,{\z2} \,{\z3} \,{\z7}
+\frac{2740777984 }{1583}\,{\z2} \,{\zp{5}{2}}
-\frac{154231727792128 }{24888291}\,{\z2} \,{\z9}\nonumber\\
&&
-24064 \,{\zp{3}{4}}
-8192 \,{\zp{3}{3}}
-790528 \,{\z4}\,{\zp{3}{2}} 
+\frac{9034356327424}{24888291}  \,{\z5}\,{\zp{3}{2}}\nonumber\\
&&
-\frac{885248 }{3} \,{\z6}\,{\zp{3}{2}}
-\frac{235947008 }{57}\,{\z3} \,{\z4} \,{\z5}
-\frac{738162401615104}{223994619} \,{\z3} \,{\z8}\nonumber\\
&&
-\frac{497325824}{211} \,{\z3} \,{\z9}
+\frac{74232684685824}{8296097} \,{\z4} \,{\z7}
-\frac{60861992014336}{24888291} \,{\z5} \,{\z6}\nonumber\\
&&
-\frac{2479374592}{1055} \,{\z5} \,{\z7}
+98304 \,{\z3} \,{\z7}
+4096 \,{\zp{5}{2}}
\Bigg) \,.
\end{eqnarray}
}
The absence of the poles in $\omega$ is a very strong test for the correctness of our seven-loop result, as the analytically continued anomalous dimension at $M=-2+\omega$ has about {\em{two hundred}} poles terms up to $g^{14}/w^2$ (with the different combinations of $\zeta_i$ and other special numbers).
Note that the generalised double-logarithmic equation~(\ref{DLgener}) can be used to control even the $\zeta_{11}$-term in the seven-loop anomalous dimension coming from eq.~(\ref{P14AnsatzZs}) as its analytic continuation is proportional to $\zeta_{11}/\omega^2$, which is not possible from the BFKL equation~(\ref{BFKLPredictions}).

\subsection{Generalised double-logarithmic equation: $\M=-r+\omega,\ r= 4,\ 6,\ \ldots$} \label{sec:GenerDL2r}
Finally, we can consider the analytic continuation of our result to the regimes around other even negative integer values, $M=-r+\omega$, with $r=4,6,\ldots$, where a generalisation of the double-logarithmic was found in ref.~\cite{Velizhanin:2011pb}. The generalisation states that around $\M=-r+\omega\,,\ r=2,4,6,\ldots$ the reciprocity-respecting function $\reciP(M)$ can be written as
\begin{equation}
{\mathcal P}_{\mathrm{DL}}(\omega,r) =2\,\sum _{k=1}\sum _{m=0}{\mathcal{D}}^k_m(r)\,\omega^m
\left(\frac{-4\,g^2}{\omega}\right)^k\,.\label{PDLr}
\end{equation}
Some of the coefficients ${\mathcal{D}}^k_m(r)$ are given in ref.~\cite{Velizhanin:2011pb}. 
To test our result, we check the fact that according to eq.~(\ref{PDLr}) $\cP_{14}(\M)$ should not contain poles higher than $1/\omega^7$, and this is indeed true.

\section{Conclusion} \label{sec:discussion}

The main result of this paper is the planar seven-loop anomalous dimension of twist-2 operators with arbitrary Lorentz spin $\M$ in $\cN=4$ SYM theory. This result, assumed to satisfy the maximal transcendentality principle, was reconstructed from a set of values at fixed spin which where found by solving the quantum spectral curve perturbatively. However, to reconstruct the rational part, we had to split the contribution into a part coming from the asymptotic Bethe ansatz plus a wrapping correction and then reconstruct these terms separately.
In some parts of the reconstruction, in particular of the wrapping part of the rational contribution, we needed to solve a system of linear equations of significantly lower rank than the number of unknowns. This was done using a special method from number theory, namely the floating point realization~\cite{fplll} of the LLL-algorithm~\cite{Lenstra:1982}.
All of these computations were done on the level of the reciprocity-respecting function $\reciP(M)$, from which the anomalous dimension can be generated using eq.~(\ref{Pfunction}). The expression for the anomalous dimension is very lengthy and therefore not written explicitly in the paper, but it is available in the ancillary files of the arXiv version of the paper and on the web-page:
\href{http://thd.pnpi.spb.ru/~velizh/7loop/}{\texttt{http://thd.pnpi.spb.ru/\textasciitilde
velizh/7loop/}}.

The obtained result was thoroughly tested against the constraints coming from the BFKL equation~(\ref{BFKLPredictions}) and the generalised double-logarithmic equations~(\ref{DLgener}), (\ref{PDLr}).
These equations provide more than {\it two hundred} constraints.
The complete agreement with these constraints confirms the correctness of the result.

One of the main aims of this work was to obtain new information about the analytic properties of the anomalous dimension of twist-2 operators. 
The result for the analytic continuation at $M=-2$ can be found in eq.~(\ref{DLgener}) and provide us with new information about the generalised double-logarithmic equation.
The result for the analytic continuation at $M=-1$, written in eq.~(\ref{BFKLPredictions}), confirm the prediction from ref.~\cite{Velizhanin:2015xsa} (and also from refs.~\cite{Gromov:2015vua} and~\cite{Caron-Huot:2016tzz}). For future calculations related with the BFKL equation, we write down the next-to-next-to-next-to-leading poles of the anomalous dimension at $M=-1$, which have the following form
\begin{eqnarray}
\gamma_{\mathrm{BFKL}}^{\mathrm{N^3LLA}}&=&
2 {\z3} \omega^3
\left(\frac{-4g^2}{\omega}\right)
-\frac{31}{4} {\z4} \omega^3
\left(\frac{-4g^2}{\omega}\right)^2
+\big(35 {\z5}
-8 {\z2} {\z3}\big) \omega^3
\left(\frac{-4g^2}{\omega}\right)^3
\nonumber\\&&
+\left(
\frac{5}{2} {\zp{3}{2}}
-\frac{349}{6} {\z6}
\right)\omega^3
\left(\frac{-4g^2}{\omega}\right)^4
+\left(
\frac{3761}{16} {\z7}
-\frac{39}{4} {\z2} {\z5}
\right)\omega^3
\left(\frac{-4g^2}{\omega}\right)^5
\nonumber\\&&
+\left(
-\frac{112}{171} {\h53}
+\frac{560}{57} {\h71}
+36 {\z2} {\zp{3}{2}}
-\frac{45575}{152} {\z3} {\z5}
-\frac{2484067}{10944} {\z8}
\right)\omega^3
\left(\frac{-4g^2}{\omega}\right)^6
\nonumber\\&&
\nonumber\\&&
+\left(
\frac{201}{32} {\z2} {\z7}
+\frac{1}{2}{\zp{3}{3}}
+\frac{1088}{3} {\z3} {\z6}
+\frac{839}{16} {\z4} {\z5}
+\frac{33719}{48} {\z9}
\right)\omega^3
\left(\frac{-4g^2}{\omega}\right)^7,\label{BFKLPredictionsNNNLLA}
\end{eqnarray}
and can be used for the test of the BFKL pomeron eigenvalue at four loops ($\mathrm{N^3LLA}$).

\acknowledgments

The authors would like to thank A. Bednyakov, L.N. Lipatov, T. \L ukowski, A. Onishchenko, A. Ray, M. Staudacher, D. Stehl\'{e} for useful discussions. We are particularly grateful to D.~Volin for collaboration on the development of the used methods and for useful comments.

The computations of roots (see Section \ref{sec:ABA}) were performed with resources provided by the North-German Supercomputing Alliance \href{https://www.hlrn.de/home/view}{(HLRN)} and by \href{https://www.cms.hu-berlin.de/de/dl/systemservice/computeservice/server/clou/standardseite}{CLOU} (Cluster of UNIX Machines at Humboldt University of Berlin).
 
The research of V.N. Velizhanin is supported by a Marie Curie International Incoming Fellowship within the 7th European Community Framework Programme, grant number PIIF-GA-2012-331484,
by DFG SFB 647 ``Raum -- Zeit -- Materie. Analytische und Geometrische Strukturen''\ and by RFBR grants 16-02-00943-a and 16-02-01143-a.
The research leading to these results has received funding from the People Programme (Marie Curie Actions) of the 7th European Community Framework Programme under REA Grant Agreement No 317089 (GATIS).




\providecommand{\href}[2]{#2}

\begingroup\raggedright

\endgroup

\end{document}